\documentstyle[preprint,eqsecnum,aps,epsfig]{revtex}
\tightenlines
\begin{document}
\draft

\title{
      Quantization of the electromagnetic field outside static black holes
      and its application to low-energy phenomena
      } 


\author{Lu\'\i s C.B. Crispino$^{1,2}$, Atsushi Higuchi$^3$, 
       and George E.A. Matsas$^1$}
\address{$^1$Instituto de F\'\i sica Te\'orica, 
         Universidade Estadual Paulista,\\
         Rua Pamplona 145, 01405-900, S\~ao Paulo, S\~ao Paulo, Brazil\\ 
         $^2$Departamento de F\'\i sica, Universidade Federal do Par\'a,\\ 
         Campus Universit\'ario do Guam\'a, 66075-900, Bel\'em, Par\'a,
         Brazil\\
         $^3$Department of Mathematics, University of York,\\ 
         Heslington, York YO10 5DD, United Kingdom}
\date{\today}
\maketitle 


\begin{abstract}

We discuss the Gupta-Bleuler quantization of the free electromagnetic 
field outside static black holes
in the Boulware vacuum.  We use a gauge which reduces to
the Feynman gauge in Minkowski spacetime. We also discuss its relation with
gauges used previously.  
Then we apply the low-energy sector of this field theory 
to investigate some low-energy phenomena. First, we discuss 
the response rate of a
static charge outside the Schwarzschild black hole in four dimensions.
Next, motivated by string physics, we compute the absorption cross 
sections of low-energy plane waves
for the Schwarzschild and extreme Reissner-Nordstr\"om black holes 
in arbitrary dimensions higher than three. 
\end{abstract} 
\pacs{04.70.Dy, 04.62.+v}

\narrowtext 

\section{Introduction}
\label{sec:Introduction}

The prediction that black holes should thermally evaporate \cite{H} 
has sparked much interest in quantum field theory in curved spacetimes.
One of the 
difficulties in studying fields in Schwarzschild~\cite{CS} 
and other black hole
spacetimes, even when the fields are non-interacting,
stems from the fact that the solutions to the field equations
are functions whose properties are not well known.\footnote{See  
Ref.~\cite{JC} for some known properties in the Schwarzschild case.}
In the low-frequency regime, however, the situation 
is much simpler in Schwarzschild spacetime.  In this regime, 
the mode functions of the massless scalar field are well known~\cite{Candel}.
The present authors and Sudarsky
used this fact to find the response rate of a static scalar source~\cite{HMS3} 
and an analytic approximation for the emission rate of low-energy particles 
from classical sources~\cite{CHM2} outside the Schwarzschild black hole in
closed form.
On the other hand, the field equations of the electromagnetic field in a
black hole spacetime are not decoupled and are difficult 
to analyze in the Lorenz gauge. However, if we require
the field to be divergence-free on a two-sphere (the spherical
Coulomb gauge), the equations for the physical modes reduce to decoupled
scalar field equations. Furthermore, solutions in terms of familiar special
functions can be found in the low-energy regime.
These observations enabled us recently to calculate
the response rate of a static electric charge outside a Schwarzschild
black hole in closed form~\cite{CHM}.  

In this paper we examine
free quantum electrodynamics in static spherically symmetric spacetimes of
arbitrary dimensions in a modified Feynman gauge.
(This gauge is closely related to the $A_0=0$ gauge 
used by Cognola and Lecca~\cite{CL} and reduces to the
Feynman gauge in Minkowski spacetime.)
Then, we calculate some low energy quantities in electrodynamics outside
spherically symmetric black holes.  First, we review the
calculation of the response rate of a static charge outside the 
four-dimensional Schwarzschild black hole in the Unruh vacuum~\cite{Unruh}. 
Next we calculate the low energy absorption cross sections of photons for
the Schwarzschild
and extreme Reissner-Nordstr\"om black holes in arbitrary 
dimensions higher than three, extending some results obtained 
by Gubser~\cite{Gubser} using a method~\cite{Teuk,Staro}
based on the Newman-Penrose formalism~\cite{NP}.

The  paper is organized as follows.
In Section~\ref{sec:GuptaBleuler} we present the mode functions of the
electromagnetic field in the spacetime of a spherically symmetric black hole
in our modified Feynman gauge.
Then we discuss the corresponding quantum theory and show how the Gupta-Bleuler
condition (see, e.g., \cite{IZ}) is implemented to obtain the physical states. 
In Section~\ref{sec:Coulomb} we compare the physical modes 
in the spherical Coulomb gauge~\cite{CHM} with the
ones obtained in the modified Feynman gauge. 
In Section~\ref{sec:Rate} we review the calculation of the response rate of
a static charge outside a four-dimensional Schwarzschild black hole.
In Sections~\ref{sec:Schwarz} and \ref{sec:Reissner} we present the
photon absorption cross sections by the Schwarzschild and 
extreme Reissner-Nordstr\"om black holes of arbitrary
dimensions higher than three.  
In Sec.~\ref{sec:Conclusions} we summarize the main results and 
make some remarks.  
In Appendix \ref{app:Mink} 
we compute some components of field-strength two-point function in 
Minkowski spacetime using
spherical polar coordinates
and show that they agree with those obtained using Cartesian coordinates.
In Appendix \ref{app:Formula} a summation formula for Legendre functions
used in Section \ref{sec:Rate}
is derived. In Appendix \ref{app:Cross}, a formula which relates the
absorption {\em probability} to the absorption {\em cross section} is
derived in arbitrary dimensions.
We use the metric signature $(+--\cdots-)$ and the natural units with
$G=\hbar=c=1$ throughout this paper.  

\section{Gupta-Bleuler quantization in a modified Feynman gauge}
\label{sec:GuptaBleuler}
In this section we analyze the field equations for the
electromagnetic field in spherically symmetric and static spacetimes in
a modified Feynman gauge. 
Then we discuss the
Gupta-Bleuler quantization in this gauge.

The line element of the spacetime we study is
\begin{equation}
d\tau^2 = f(r)dt^2 - h(r)dr^2 - r^2 ds^2_p\,,
\label{ssle}
\end{equation}
where $ds^2_p$ is the line element of a unit $p$-sphere.  We assume that
$f(r)$ and $h(r)$ are positive for $r >r_H$ and
that both $f(r)$ and $h(r)^{-1}$ have simple zeroes or both have double zeroes 
at $r=r_H$, where $r_H$ is the horizon radius. 
We also assume that $f(r), h(r)^{-1} \to 1$ as $r\to \infty$.
(Most results in this section, however, are independent of these assumptions.)
Let us introduce the Wheeler tortoise coordinate $r^*$ by
\begin{equation}
\frac{dr^*}{dr} = \sqrt{\frac{h}{f}}\,. \label{tortoise}
\end{equation}
Then $r^*(r)$ is a monotonic function with domain 
$(r_H, +\infty)$ and range in $(-\infty,+\infty)$. Let us define for
any two functions $q_1(r^*)$ and $q_2(r^*)$ the inner product
\begin{equation}
\langle q_1,q_2\rangle = 
\int_{-\infty}^{+\infty}dr^* \overline{q_1(r^*)}q_2(r^*)\,, \label{2dinner}
\end{equation}
where the overline denotes complex conjugation.

The Lagrangian density of the electromagnetic field in a modified Feynman
gauge is 
\begin{equation}
{\cal L}_{\rm F} = \sqrt{-g}\left[ -\frac{1}{4}F_{\mu\nu}F^{\mu\nu}
- \frac{1}{2}G^2\right]\,
\end{equation}
with
\begin{equation}
G = \nabla^\mu A_\mu + K^\mu A_\mu\,,  \label{eqforG}
\end{equation}
where the vector $K^\mu$ is independent of $A_\mu$. 
Hence the equations of motion are
\begin{equation}
\nabla_\nu F^{\nu\mu} +\nabla^\mu G - K^\mu G = 0\,. \label{geneq}
\end{equation}
Here we choose
\begin{equation}
K^\mu = ( 0, f'/(fh),0,0)\,, \label{Kchoice}
\end{equation}
in which case Eq.~(\ref{eqforG}) is written as
\begin{equation}
G = \frac{1}{f}\partial_t A_t - \sqrt{\frac{f}{h}}\frac{1}{r^p}
\partial_r\left[ \frac{r^p}{\sqrt{fh}}A_r\right] 
- \frac{1}{r^2}\tilde{\nabla}^i A_i\,.
\label{G}
\end{equation}
Here $i$ denotes angular variables on the unit $p$-sphere $S^p$
with metric $\tilde{\eta}_{ij}$ and inverse metric $\tilde{\eta}^{ij}$ 
[with signature $(+\cdots+)]$, $\tilde{\nabla}_i$
is the associated covariant derivative on $S^p$ and  
$\tilde{\nabla}^i \equiv \tilde{\eta}^{ij}\tilde{\nabla}_j$. 
This choice for $K^\mu$ is convenient because the equation for $A_t$ 
decouples from the other ones. The field equations~(\ref{geneq}) become 
\begin{eqnarray}
& & -\frac{1}{f} \partial_t^2 A_t + \sqrt{\frac{f}{h}\,} \frac{1}{r^p}
\partial_r \left[ \frac{r^p }{\sqrt{fh\,}}\partial_r A_t \right] 
+ \frac{1}{r^2}\tilde \nabla^2 A_t = 0 \,,
\label{eqt} \\
& & -\frac{1}{f} \partial_t^2 A_r + \frac{1}{f} \partial_r \left[ 
\sqrt{\frac{f}{h}} \frac{f}{r^p} \partial_r \left( \frac{r^p }{\sqrt{fh\,}} A_r
\right)\right]+ \frac{1}{r^2}\tilde \nabla^2 A_r
+ \frac {1}{f}\partial_r \left( \frac{f}{r^2} \right) \tilde \nabla^i A_i =0\,,
\label{eqr} \\
& & -\frac{1}{f} \partial_t^2 A_i + \frac{r^{2-p}}{\sqrt{fh\,}} 
\partial_r\left( \sqrt{\frac{f}{h}} r^{p-2} \partial_r A_i\right)
- \frac{r^2}{fh} 
\partial_r\left(\frac{f}{r^2}  \right) \partial_i A_r 
+\frac{1}{r^2} \left[ \tilde \nabla ^j (\tilde\nabla_j A_i-\tilde\nabla_i A_j)
 \right.
\nonumber \\
& & \left.
+ \partial_i(\tilde\nabla^j A_j)\right] = 0\,,
\label{eqang}
\end{eqnarray}
where
$\tilde{\nabla}^2 \equiv \tilde{\eta}_{ij} \tilde\nabla^i\tilde\nabla^j$.

We shall now
describe a complete set of solutions $A_\mu^{(\lambda n;\, \omega lm)}$.
We assign $\lambda$ the value 0 for what we call the non-physical modes,
1 or 2 for the physical modes and 
3 for the pure-gauge modes. (These modes will be given below.) 
The label $n$ distinguishes between modes incoming from the past null
infinity ${\cal J}^-$ (denoted with $n = \leftarrow$) 
and those coming out from the past horizon $H^{-}$ (denoted with 
$n = \rightarrow$).\footnote{Here
we treat only the solutions proportional to $e^{-i\omega t}$ with
$\omega \neq 0$.  Thus, if there is any nonzero static field, we will be
considering fluctuation about that solution.}
The solutions  with $A_\mu^{(0n;\,\omega lm)} = 0$, $(\mu \neq t)$, and
\begin{equation}
A_t^{(0n;\,\omega l m)} = R_{\omega l}^{(0n)}(r)Y_{lm}e^{-i\omega t}
\end{equation}
will be called ``non-physical modes'' because they satisfy the 
field equations~(\ref{eqt})-(\ref{eqang})
but not the gauge condition $G=0$. Here
$Y_{lm}$ is a scalar spherical harmonic on the unit $p$-sphere with
$\tilde{\nabla}^2 Y_{lm} = -l(l+p-1)Y_{lm}$, where $l=0,1,2,\ldots$
and $m$ denotes a set of $p-1$ integers $(m_1,\ldots,m_{p-1})$
satisfying $l\ge m_{p-1}\ge \ldots \ge m_{2} \ge |m_1|$.
(See Ref.~\cite{CM} for a concise description of spherical 
harmonics on the $p$-sphere.) They are normalized as
\begin{equation}
\int d\Omega_p \overline{Y_{lm}}Y_{l'm'} = \delta_{ll'}\delta_{mm'}\,,
\end{equation}
where $d\Omega_p$ is the volume element of the unit $p$-sphere.
The function $R_{\omega l}^{(0n)}(r)$ satisfies
\begin{equation}
\left[ \frac{\omega^2}{f} 
+ \sqrt{\frac{f}{h}}\frac{1}{r^p}\frac{d}{dr}
\left( \frac{r^p}{\sqrt{fh}} \frac{d}{dr}
\right) - \frac{l(l+p-1)}{r^2}\right]R_{\omega l}^{(0n)}= 0\,.  \label{eqrA_t}
\end{equation}
We will determine the normalization of the functions $R_{\omega l}^{(0n)}(r)$
later.  
The  pure-gauge modes are given as
\begin{equation}
A_\mu^{(3n;\,\omega lm)}=\nabla_{\mu}\Lambda^{(n\omega lm)}\,,
\end{equation}
where
\begin{equation}
\Lambda^{(n\omega lm)}= \frac{i}{\omega}
R_{\omega l}^{(0n)}(r)Y_{lm}e^{-i\omega t}
\,.
\end{equation}
(As usual, the pure-gauge modes satisfy the field 
equations~(\ref{eqt})-(\ref{eqang}) and $G=0$.)
The other independent solutions $(\lambda=1,2)$, which represent 
physical degrees of freedom, i.e., which satisfy the field equations and
$G=0$ but are not pure gauge, will be chosen to have 
$A_t = 0$.
Those with $A_r \neq 0$
are given as
\begin{equation}
A_r^{(1n;\,\omega lm)} = R_{\omega l}^{(1n)}(r)Y_{lm}e^{-i\omega t}\,,
\end{equation}
where
\begin{equation}
\left[ \frac{\omega^2}{f}-\frac{l(l+p-1)}{r^2}\right] R_{\omega l}^{(1n)}(r)
+ \frac{1}{r^2}\frac{d}{dr}
\left[ \sqrt{\frac{f}{h}\,}r^{2-p}\frac{d}{dr} \left(
\frac{r^p}{\sqrt{fh}} R_{\omega l}^{(1n)}(r) \right)\right]
= 0\,.
\label{R^1}
\end{equation}
Note here that the condition $G=0$ cannot be solved if $l=0$.  Hence,
we have $l \geq 1$.
The corresponding angular components can be found by solving the condition
$G=0$ as
\begin{equation}
A_i^{(1n;\,\omega lm)} = \frac{r^{2-p}}{l(l+p-1)}\sqrt{\frac{f}{h}}
\frac{d}{dr} \left[ \frac{r^p}{\sqrt{fh\,}}R_{\omega l}^{(1n)}
\right]\partial_i Y_{lm}\; e^{-i\omega t} \,.
\end{equation}
We call these modes ``physical modes I".
The other set of physical solutions can be obtained by letting 
$A_t= A_r =0$ and
\begin{equation}
A_i^{(2n;\,\omega lm)} = R_{\omega l}^{(2n)}(r)Y_i^{(lm)}e^{-i\omega t}\,,
\end{equation}
where
\begin{equation}
\left[ \frac{\omega^2}{f} + \frac{1}{\sqrt{fh\,}r^{p-2}}\frac{d}{dr}
\left( \sqrt{\frac{f}{h}\,} r^{p-2}\frac{d}{dr} \right)
- \frac{(l+1)(l+p-2)}{r^2}\right]R_{\omega l}^{(2n)}(r) = 0\,.
\label{R^2}
\end{equation}
Here, the $Y_i^{(lm)}$ are divergence-free vector 
spherical harmonics on the unit $p$-sphere
satisfying 
\begin{equation}
\tilde{\nabla}^k (\tilde{\nabla}_k Y_i^{(lm)} - \tilde{\nabla}_i Y_k^{(lm)})
= - (l+1)(l+p-2)Y_i^{(lm)}
\end{equation}
and
\begin{equation}
\int d\Omega_p \tilde{\eta}^{ij}\overline{Y_i^{(lm)}}Y_j^{(l'm')} = 
\delta_{ll'}\delta_{mm'}\,.
\end{equation}
(See, e.g., Refs.~\cite{CM,CH}.)  We call these modes ``physical modes II".
Physical modes I and II obtained here (and restricted to four
dimensions) are identical with those in the $A_0=0$ gauge~\cite{CL}.

In order to discuss Gupta-Bleuler
quantization of this field it is convenient to introduce a
generalized Klein-Gordon product of classical solutions of Eq.\ (\ref{geneq}).
We first define 
\begin{equation}
\Pi^{\mu\nu} \equiv
\frac{1}{\sqrt{-g}}\frac{\partial{\cal L}_{\rm F}}
{\partial[\nabla_\mu A_\nu]}
= -\left[F^{\mu\nu} + g^{\mu\nu} G
\right]\,,
\end{equation}
where $G$ is given in Eq.~(\ref{eqforG}).
Note that $\sqrt{-g}\,\Pi^{t\nu}$ is the canonical conjugate momentum of 
$A_\nu$.  We write
\begin{equation} 
\Pi^{(\zeta)\mu\nu} \equiv
\left. \Pi^{\mu\nu}\right|_{A_\mu = A_\mu^{(\zeta)}}
\end{equation}
for any solution $A_\mu^{(\zeta)}\equiv A_\mu^{(\lambda n ;\, \omega lm)}$.  
For any two (complex) solutions
$A_\mu^{(\zeta)}$ and $A_\mu^{(\zeta')}$ we define
\begin{equation}
W^\mu[A^{(\zeta)},A^{(\zeta')}]
\equiv i \left[\overline{A_\nu^{(\zeta)}}\Pi^{(\zeta')\mu\nu}
- \overline{\Pi^{(\zeta)\mu\nu}}A_\nu^{(\zeta')}\right]\,.
\end{equation}
The field equations ensure that this current is
conserved. 
As a result, the generalized Klein-Gordon inner product defined by  
\begin{equation}
(A^{(\zeta)},A^{(\zeta')}) \equiv \int_\Sigma d\Sigma_\mu
W^\mu[ A^{(\zeta)},A^{(\zeta')}]\,, \label{KGinner}
\end{equation}
where $d\Sigma_\mu \equiv d\sigma\, n_\mu$, is independent of the 
Cauchy surface
$\Sigma$~\cite{Fri}. (Here $d\sigma$ is the
volume element of the Cauchy surface $\Sigma$ with a normal unit vector
$n^\mu$.)  Note that 
$(A^{(\zeta)},A^{(\zeta')}) =  \overline{(A^{(\zeta')},A^{(\zeta)})}$.
This guarantees that the norm defined through (\ref{KGinner}) is real (and
positive definite for the subset of physical solutions with
positive frequency).
By working explicitly with the definition (\ref{KGinner}), we find in 
general that on a $t={\rm const}$ surface
\begin{equation}
\left( A^{(\zeta)}, A^{(\zeta')}\right)
= - i \int d^{p+1}{\bf x} \sqrt{-g}f^{-1}g^{\mu\nu}
\left( \overline{A_\mu^{(\zeta)}}\partial_t A_\nu^{(\zeta')}
- A_\nu^{(\zeta')} \partial_t \overline{A_\mu^{(\zeta)}} \right)\,.
\label{explicit}
\end{equation}

It is important to note
that pure-gauge modes are orthogonal to any mode satisfying $G=0$ and,
as a result, $\nabla_\nu F^{\nu \mu} = 0$.
This can be shown as follows.   Suppose that $A^{(\zeta)}_\mu$ and
$A^{(\zeta')}_\mu$ satisfy the condition $G=0$.  Then  since
$\Pi^{(\zeta)\mu\nu} =-\left.F^{\mu\nu}\right|_{A_\mu = A_\mu^{(\zeta)}}$,
and similarly for $A_\mu^{(\zeta')}$,
the inner product (\ref{KGinner}) can be written as
$(A^{(\zeta)}, A^{(\zeta')})=(A^{(\zeta)}, A^{(\zeta')})_{\rm inv}$
where
\begin{equation}
(A^{(\zeta)}, A^{(\zeta')})_{\rm inv}
\equiv i \int_\Sigma d\Sigma_\mu \left[ \overline{A_\nu^{(\zeta)}}
F^{(\zeta')\nu\mu}
- \overline{F^{(\zeta)\nu\mu}}A_\nu^{(\zeta')}\right]\,.  \label{Newinner}
\end{equation}

Now, let $A_\mu^{(\zeta)} = \nabla_\mu \Lambda^{(\zeta)}$ be a pure-gauge mode.
Then
\begin{equation}
(A^{(\zeta)}, A^{(\zeta')})_{\rm inv}
= i \int_\Sigma d\Sigma_\mu 
F^{(\zeta')\nu\mu}\nabla_\nu\overline{\Lambda^{(\zeta)}}
= i \int_\Sigma d\Sigma_\mu \nabla_\nu \left( \overline{\Lambda^{(\zeta)}}\;
F^{(\zeta')\nu\mu}\right) = 0  \label{invar}
\end{equation}
since $F^{(\zeta')\nu\mu}$ is anti-symmetric.\footnote{Note that
Eq.\ (\ref{invar}) would hold even if $\Lambda^{(\zeta)}$ was an arbitrary
function.  This shows that 
$(A^{(\zeta)}, A^{(\zeta')})_{\rm inv}$
is gauge invariant.} Thus, 
$(A^{(\zeta)},A^{(\zeta')}) = 0$ if $A^{(\zeta)}_\mu$ is a pure-gauge
mode and $A^{(\zeta')}$ is a physical or pure-gauge mode, i.e.,
\begin{equation}
(A^{(3n;\,\omega lm)}, A^{(3n';\,\omega' l'm')}) = 0\,,
\label{A3A3}
\end{equation}
\begin{equation}
(A^{(3n;\,\omega lm)}, A^{(1n';\,\omega' l'm')}) = 0\;, \;
(A^{(3n;\,\omega lm)}, A^{(2n';\,\omega' l'm')}) = 0\,.
\label{ApA3}
\end{equation}
Next let us examine the inner product of the
non-physical solutions.
By letting
\begin{equation}
R^{(0n)}_{\omega l}(r) \equiv \sqrt{f}\; r^{-p/2}\varphi^{(0n)}_{\omega l}(r)\,,
\end{equation}
we find  from Eq.~(\ref{eqrA_t})
\begin{equation}
\left( \omega^2 + \frac{d^2}{{dr^*}^2} - V_0(r^*) \right)
\varphi^{(0n)}_{\omega l} = 0\,,
\label{fi0}
\end{equation}
where $r^*$ is given in Eq.\ (\ref{tortoise}) and
$$
V_0[r^*(r)] = 
 f\frac{l(l+p-1)}{r^2}
-\frac{f}{h}\left[\frac{p(2-p)}{4r^2} - \frac{f^{\prime 2}}{2f^2}
+ \frac{p}{4r}\left( \frac{f'}{f} + \frac{h'}{h}\right) + \frac{f''}{2f}
- \frac{f'h'}{4fh}\right] \,.
$$
For spacetimes where $f(r)$ and $h(r)^{-1}$ have simple zeroes at $r=r_H$,
we find $V_0 [r^*(r_H)] \neq 0$ in general.  If $V_0[r^*(r_H)] > 0$, then 
the non-physical modes coming out from the horizon have 
frequencies satisfying $\omega^2 > V_0[r^*(r_H)]$. In particular, for  
Reissner-Nordstr\"om spacetime in (p+2)-dimensions~\cite{MP}, where
$$
f(r) = h(r)^{-1} =\left[ 1- \left(\frac{r_+}{r}\right)^{p-1}\right]
\left[ 1- \left(\frac{r_-}{r}\right)^{p-1}\right]
$$
with $r_\pm^{p-1} = M \pm \sqrt{M^2-Q^2\,}$ (so that $r_H = r_+$), we find  
$$
{V_0 [r^*(r_H)] } = \frac{(p-1)^2}{(2r_+)^2}
\left[ 1- \left(\frac{r_-}{r_+}\right)^{p-1}\right]^2
\,.
$$
As expected, we have 
$\langle \varphi^{(0\rightarrow)}_{\omega l},
\varphi^{(0\leftarrow)}_{\omega' l}\rangle = 0$, i.e.
the solutions of Eq.~(\ref{fi0})
incoming from $H^{-}$ and those incoming
from ${\cal  J}^-$
are orthogonal to one another with respect to the inner
product defined by 
Eq.\ (\ref{2dinner}). 
By normalizing these solutions so that
\begin{eqnarray}
\varphi^{(0\rightarrow)}_{\omega l} & \approx &
\sqrt{{\omega}/{\tilde\omega}\,}
\left( e^{i\tilde\omega r^*} + 
{\cal R}^{(0\rightarrow)}_{\omega l} e^{-i\tilde\omega r^*} \right)
\,\ \ \ (r^* \to -\infty)\,, \\
\varphi^{(0\leftarrow)}_{\omega l} & \approx &
e^{-i\omega r^*} + {\cal R}^{(0\leftarrow)}_{\omega l} e^{i\omega r^*}
\,\ \ \ (r^* \to +\infty)\,, 
\end{eqnarray}
up to a phase factor, where $\tilde\omega^2 = 
\omega^2-V_0[r^*(r_H)] \geq 0$ and
${\cal R}^{(0\rightarrow)}_{\omega l}$ and 
${\cal R}^{(0\leftarrow)}_{\omega l}$ 
are constants (with 
$|{\cal R}^{(0\rightarrow)}_{\omega l}|
=
|{\cal R}^{(0\leftarrow)}_{\omega l}|$), we have
\begin{equation}
\langle \varphi^{(0\leftarrow)}_{\omega l},
\varphi^{(0\leftarrow)}_{\omega' l}\rangle = 
\langle \varphi^{(0\rightarrow)}_{\omega l},
\varphi^{(0\rightarrow)}_{\omega' l}\rangle
= 2\pi \delta(\omega-\omega') \,.
\end{equation}
Now, by using Eq.\ (\ref{explicit}) we find
\begin{eqnarray}
(A^{(0n;\,\omega lm)}, A^{(0n';\,\omega' l'm')})
 & = & - 2\omega \langle \varphi^{(0n)}_{\omega l},\varphi^{(0n')}_{\omega' l'}
 \rangle
\delta_{nn'}\delta_{ll'}\delta_{mm'}
\nonumber \\
& = & 
- 4\pi \omega \delta_{nn'}\delta_{ll'}\delta_{mm'}\delta(\omega -\omega')\,.
\end{eqnarray}
where we have used 
$
\delta(\tilde\omega -\tilde\omega') 
= 
(\tilde\omega/\omega)\delta(\omega -\omega')\,.
$
Noting that $A^{(0n;\,\omega lm)}$ and $A^{(3n;\,\omega lm)}$
have the same $t$-components, 
we immediately obtain from Eqs.\ (\ref{explicit}) 
and (\ref{A3A3}) that
\begin{equation}
(A^{(0n;\,\omega lm)}, A^{(3n';\,\omega' l'm')})
= - 4\pi \omega \delta_{nn'}\delta_{ll'}\delta_{mm'}\delta(\omega -\omega')\,.
\end{equation}

Next let us examine the physical modes.  By letting 
\begin{equation}
R^{(1n)}_{\omega l}(r) \equiv
\frac{\sqrt{l(l+p-1)}}{\omega}\, (fh)^{1/2} r^{-p/2-1} 
\varphi^{(1n)}_{\omega l}(r)\,,
\label{R1}
\end{equation}
we find from Eq.~(\ref{R^1})
\begin{equation}
\left( \omega^2 + \frac{d^2\ }{dr^{* 2}} - 
V_1(r^*) \right)\varphi^{(1n)}_{\omega l}
= 0\,,  \label{varphi1}
\end{equation}
where
\begin{equation}
V_1 [r^*(r)]= f\frac{l(l+p-1)}{r^2} + \frac{p(p-2)}{4r^2}\frac{f}{h}
- \frac{(p-2)}{4} \frac{f}{h r} \left(\frac{f'}{f} - \frac{h'}{h}\right)\,.
\end{equation}
(We note that $V_1 \to 0$ as $r\to r_+$ for Reissner-Nordstr\"om spacetime
unlike $V_0$.)
We first note that the modes $A^{(1n;\,\omega lm)}$ are orthogonal to
both the non-physical and pure-gauge modes.  Next we note
\begin{equation}
(A^{(1n;\,\omega lm)}, A^{(1n';\,\omega' l'm')})
= 2\omega \langle \varphi^{(1n)}_{\omega l},
\varphi^{(1n')}_{\omega' l'}\rangle
\delta_{nn'}\delta_{ll'}\delta_{mm'}\,. \label{normal1}
\end{equation}
Thus, by normalizing $\varphi^{(1n)}_{\omega l}(r)$ as 
\begin{eqnarray}
\varphi^{(1\rightarrow)}_{\omega l} & \approx &
 e^{i \omega r^*} + 
{\cal R}^{(1\rightarrow)}_{\omega l} e^{-i \omega r^*} 
\,\ \ \ (r^* \to -\infty)\,, \label{norm1}\\
\varphi^{(1\leftarrow)}_{\omega l} & \approx &
e^{-i\omega r^*} + {\cal R}^{(1\leftarrow)}_{\omega l} e^{i\omega r^*}
\,\ \ \ (r^* \to +\infty)\,, \label{norm2}
\end{eqnarray}
we have
\begin{equation}
(A^{(1n;\,\omega lm)}, A^{(1n';\,\omega' l'm')})
= 4\pi\omega \delta_{nn'}\delta_{ll'}\delta_{mm'}
\delta(\omega - \omega')\,.
\end{equation}

Finally, we let 
$$
R^{(2n)}_{\omega l}(r) \equiv r^{-(p-2)/2}\varphi^{(2n)}_{\omega l}(r) \,.
$$  Then from Eq.~(\ref{R^2}), we have
\begin{equation}
\left( \omega^2 + \frac{d^2\ }{dr^{* 2}} - V_2(r^*) 
\right)\varphi^{(2n)}_{\omega l}
= 0\,, \label{varphi2}
\end{equation}
where
\begin{equation}
V_2 [r^*(r)]= f\frac{(l+1)(l+p-2)}{r^2} + \frac{(p-2)(p-4)}{4r^2}\frac{f}{h}
+ \frac{(p-2)}{4} \frac{f}{hr}\left(\frac{f'}{f} - \frac{h'}{h}\right)\,.
\end{equation}
(We again note that $V_2 \to 0$ as $r\to r_+$ in Reissner-Nordstr\"om
spacetime.)  Notice that $V_1 = V_2$
if $p=2$.  Hence, in the four-dimensional case 
$\varphi^{(1n)}_{\omega l}(r)$ and $\varphi^{(2n)}_{\omega l}(r)$ satisfy
the same equation.
These modes can easily be shown to be orthogonal to the modes previously 
discussed. 
By normalizing the functions $\varphi^{(2n)}_{\omega l}(r)$
in the same way as 
$\varphi^{(1n)}_{\omega l}(r)$  
[see Eqs.~(\ref{norm1})-(\ref{norm2})], we have
\begin{equation}
(A^{(2n;\,\omega lm)}, A^{(2n';\,\omega' l'm')})
= 4\pi\omega \delta_{nn'}\delta_{ll'}\delta_{mm'}
\delta(\omega - \omega')\,.
\end{equation}

In order to 
quantize the field $A_\mu$, we impose the equal-time commutation relations
on the field $\hat{A}_\mu$ and momentum $\hat{\Pi}^{t \mu }$ operators:
\begin{equation}
\left[ \hat{A}_\mu(t,{\bf x}), \hat{A}_\nu(t,{\bf x}')\right]
= \left[ \hat{\Pi}^{t\mu}(t,{\bf x}), \hat{\Pi}^{t\nu}(t,{\bf x}')\right]= 0\,,
\label{CR1}
\end{equation}
\begin{equation}
\left[ \hat{A}_\mu(t,{\bf x}), \hat{\Pi}^{t\nu}(t,{\bf x}')\right] = 
\frac{i\delta_\mu^\nu}{\sqrt{-g}}\delta^{p+1}({\bf x}-{\bf x}')\,,
\label{CR2}
\end{equation}
where ${\bf x}$ and ${\bf x}'$ represent all spatial coordinates.
The field $\hat{A}_\mu$ can be expanded using the modes we have 
obtained before: 
\begin{equation}
\hat{A}_{\mu}(t,{\bf x})
= \sum_{\rho} \int_{-\infty}^{+\infty}
\,\frac{d\omega}{\sqrt{4\pi|\omega|}} 
A_\mu^{(\omega\rho)}(t,{\bf x})a_{\omega\rho}\,,
\end{equation}
where
$A_\mu^{(\omega\rho)}$ is proportional to $e^{-i\omega t}$,
$A_\mu^{(-\omega \rho )} \equiv \overline{A_\mu^{(\omega \rho)}}$,
$a_{-\omega\rho} \equiv a^\dagger_{\omega\rho}$
and  $\rho$ labels discrete quantum numbers.  
The commutation relations~(\ref{CR1})-(\ref{CR2}) are equivalent 
to the following commutation relations in the 
``symplectically smeared" form as is the case for scalar fields~\cite{Wald}:
\begin{equation}
\left[ (A^{(\zeta)}, \hat{A}), (\hat{A}, A^{(\zeta')})\right]
= (A^{(\zeta)}, A^{(\zeta')})\,.  \label{smeared}
\end{equation} 
Since the inner product must be $t$ independent, the inner product of
$A_\mu^{(\omega\rho)}$ and $A_\mu^{(\omega'\rho')}$ can be nonzero only
if $\omega  = \omega'$.  Thus, we can write
\begin{equation}
(A^{(\omega\rho)}, A^{(\omega'\rho')})
= M^{\rho\rho'}\delta(\omega-\omega')\,. \label{Mdef}
\end{equation}
By using Eq.~(\ref{Mdef}) in (\ref{smeared}) one finds
\begin{equation}
M^{\rho_1\rho_2}\left[ a_{\omega\rho_2},a^\dagger_{\omega'\rho_3}\right]
M^{\rho_3\rho_4} = 4\pi\omega M^{\rho_1\rho_4}\delta(\omega-\omega')\,.
\end{equation}
Then, since $M^{\rho\rho'}$ is invertible in our case, we find
\begin{equation}
\left[ a_{\omega\rho},a^\dagger_{\omega' \rho'}\right] = 4\pi\omega
(M^{-1})_{\rho\rho'}\delta(\omega-\omega')\,. \label{acommu}
\end{equation}
Note here that we immediately have $[a_{\omega \rho},a_{\omega' \rho'}] = 0$
for $\omega, \omega' > 0$ by letting $\omega' \to -\omega'$ 
in Eq.~(\ref{acommu}) .
By using the Klein-Gordon inner products computed above, we find the following
commutators:
\begin{eqnarray}
 \left[ a^{(3n)}_{\omega lm}, a^{(3n')\dagger}_{\omega' l'm'}\right]
= -\left[ a^{(0n)}_{\omega lm}, a^{(3n')\dagger}_{\omega' l'm'}\right]
&=& \delta_{nn'}\delta_{ll'}\delta_{mm'}\delta(\omega-\omega')\,, \\
\left[ a^{(1n)}_{\omega lm}, a^{(1n')\dagger}_{\omega' l'm'}\right]
= \left[ a^{(2n)}_{\omega lm}, a^{(2n')\dagger}_{\omega' l'm'}\right]
& = & \delta_{nn'}\delta_{ll'}\delta_{mm'}\delta(\omega-\omega')
\end{eqnarray}
with all other commutators vanishing.
The Gupta-Bleuler condition~\cite{IZ}
requires that any physical state $|{\rm phys}\rangle$
satisfy
\begin{equation}
\hat{G}^{(+)}|{\rm phys}\rangle = 0\,,
\end{equation}
where $\hat{G}^{(+)}$ is the positive-frequency part of 
$\hat{G} = \nabla^\mu \hat{A}_\mu + K^\mu \hat{A}_\mu$.  Since this quantity is
nonvanishing only for $A^{(0n;\,\omega lm)}_\mu$, this
condition is equivalent to
\begin{equation}
a^{(0n)}_{\omega l m}|{\rm phys}\rangle = 0\ \ \ {\rm for\ all\ }
(n,\omega,l, m)\,\ ({\rm with}\ \omega > 0)\,. 
\end{equation}
(We let $\omega > 0$ below.)
The Boulware vacuum $|0\rangle$~\cite{Boul}
is defined by requiring that it be annihilated
by all $a^{(\lambda n)}_{\omega lm}$ operators  
$(\lambda=0,1,2,3)$.
Note that the states obtained by applying any number of creation operators
excluding $a^{(3n)\dagger}_{\omega lm}$ are all physical states.  Any
state of the form $a^{(3n)\dagger}_{\omega lm}|{\rm phys}\rangle$ is unphysical
because
$$
a^{(0n')}_{\omega'l'm'}a^{(3n)\dagger}_{\omega lm}|{\rm phys}\rangle
= - \delta_{nn'}\delta_{ll'}\delta_{mm'}\delta(\omega-\omega')
|{\rm phys}\rangle \neq 0\,.
$$
Note also
that the physical states of the form
$a^{(0n)\dagger}_{\omega lm}|{\rm phys}\rangle$ have zero norm and are
orthogonal to any physical states.  Thus, as is well known, 
a physical state $|{\rm phys}_1\rangle$ can be regarded as equivalent to
any state of the form 
$|{\rm phys}_1\rangle + a^{(0n)\dagger}_{\omega lm}|{\rm phys}_2\rangle$.
We can take as the representative elements the states obtained by applying 
$a^{(\lambda n)\dagger}_{\omega lm}$, $\lambda = 1,2$, on $|0\rangle$. 

Unphysical particles created by $a^{(3n)\dagger}_{\omega lm}$ will be in 
thermal equilibrium in the Hartle-Hawking vacuum~\cite{HH}
for a static black hole
if we require the gauge-fixed two-point function be non-singular on the
horizons as in the scalar case~\cite{KW}.  There will also be a flux of
unphysical particles in the Unruh vacuum.  Therefore, technically speaking,
the Hartle-Hawking
and Unruh vacua
are unphysical if we impose the Gupta-Bleuler condition using the 
positive-frequency notion in the Boulware vacuum.  Fortunately, 
the BRST quantization does not suffer from this problem
since the physical state
condition does not refer to any notion of positive frequency~\cite{KO}.
This issue, however, will not concern us in the following, since all our 
calculations will be at the tree level. 

Mode expansions in spherical polar coordinates are not very widely used and are 
quite different from the ordinary method using Cartesian coordinates.
For this reason, we compute in Appendix A some components 
of field-strength two-point
function in four-dimensional Minkowski spacetime using spherical
polar coordinates.  
We find that they agree
with the standard results obtained in Cartesian coordinates, as they should.

\section{The physical modes in the spherical Coulomb gauge}
\label{sec:Coulomb}
As we have seen, the modified Feynman gauge defined by Eq.~(\ref{Kchoice}) 
is useful in finding explicit
mode functions.  We also noted that this gauge results in the same physical
modes as in the $A_0=0$ gauge.
There is another convenient gauge for finding physical 
modes, i.e., the spherical Coulomb gauge, $\tilde{\nabla}^i A_i = 0$.
(One can readily show that any vector $A_\mu$ is gauge-equivalent
to a vector satisfying this gauge condition.)
Let us discuss the relation
of the physical modes obtained in this gauge and those found in the previous
section.
\footnote{Again, 
we only treat solutions proportional to $e^{-i\omega t}$ with
$\omega \neq 0$.}

It is clear that the physical modes $A^{(2n;\,\omega lm)}_\mu$ in the
previous section are also in the spherical Coulomb gauge.  The other independent
physical modes must have $A_i = 0$ in this gauge.  
First we note that if $A_t$ and $A_r$ are assumed to be spherically 
symmetric, then we find from the equations
$\nabla^\mu F_{\mu t} = \nabla^\mu F_{\mu r} = 0$
that $F_{tr}$ is $t$-independent and $(r^p/\sqrt{fh})F_{tr}$ is 
$r$-independent.  Hence, if in addition the modes are assumed to be 
proportional to $e^{-i\omega t}$ with $\omega \neq 0$, they must be 
pure gauge because $F_{tr} = 0$. (A time-independent solution has
$F_{tr} \propto\sqrt{fh}/r^p$. This
will represent a static Coulomb field, which we will disregard here.) 
To find the solutions which are not spherically symmetric we let
$A_t, A_r \propto Y_{lm}e^{-i\omega t}$. From the equation
$\nabla^\mu F_{\mu i}=0$ we find  
\begin{equation}
-i\omega A_t = \sqrt{\frac{f}{h}}\frac{1}{r^{p-2}}
\partial_r \left[ \sqrt{\frac{f}{h}} r^{p-2}A_r\right]\,. \label{Atmode}
\end{equation}
By substituting this and $A_i =0$ in $\nabla^\mu F_{\mu r} = 0$, we find
that $A_r^{(1'n;\,\omega l m)}= R^{(1'n)}_{\omega l}(r)Y_{lm}e^{-i\omega t}$
(here we use primed indices to refer to the spherical Coulomb gauge), where
\begin{equation}
\left[ \frac{\omega^2}{f} -\frac{l(l+p-1)}{r^2}\right] R^{(1'n)}_{\omega l}(r)
+ \frac{1}{f}\frac{d}{dr}\left[ \sqrt{\frac{f}{h}}\frac{1}{r^{p-2}}
\frac{d}{dr} \left( 
\sqrt{\frac{f}{h}}r^{p-2}R^{(1'n)}_{\omega l}(r)\right)\right]
= 0\,.
\end{equation}
The component $A_t^{(1'n;\,\omega lm)}$ can be found from Eq.~(\ref{Atmode}).
(Then, one can readily show that $\nabla^\mu F_{\mu t}=0$.)
Indeed one can verify that these modes are related to the ones 
in the modified Feynman gauge by a gauge transformation:
\begin{equation}
A_\mu^{(1 n;\,\omega lm)} \to A_\mu^{(1'n;\,\omega lm)}
= A_\mu^{(1 n;\,\omega lm)} - \nabla_\mu \Phi^{(1 n;\,\omega lm)}\,,
\end{equation}
where
\begin{equation}
\Phi^{(1 n;\,\omega lm)} = \frac{r^{2-p}}{l(l+p-1)}
\sqrt{\frac{f}{h}}\frac{d}{dr} \left( \frac{r^p}{\sqrt{fh}}R^{(1n)}_{\omega l}
\right) Y_{lm} e^{-i\omega t}\,.
\end{equation}
This gives us
\begin{eqnarray}
A_t^{(1'n;\,\omega lm)}
& = & \frac{i\omega r^{2-p}}{l(l+p-1)}\sqrt{\frac{f}{h}}\frac{d}{dr}
\left( \frac{r^p}{\sqrt{fh}}R^{(1n)}_{\omega l}\right)Y_{lm}
e^{-i\omega t}\,, \label{AtCoul} \\
A_r^{(1'n;\,\omega lm)}
& = & \frac{\omega^2 r^{2}}{l(l+p-1)}\frac{1}{f}R^{(1n)}_{\omega l}Y_{lm}
e^{-i\omega t}\,. \label{suppr}
\end{eqnarray}

\section{Response rate of a static charge outside a four-dimensional
Schwarzschild black hole}
\label{sec:Rate}

Here we briefly review, in the context of the previous two sections, 
the calculation of response rate in the Unruh vacuum~\cite{Unruh}
of a static electric charge in Schwarzschild spacetime
performed in Ref.~\cite{CHM}.

The line element of the four-dimensional Schwarzschild spacetime is
\begin{equation}
d\tau^2 = f(r)dt^2 - f(r)^{-1}dr^2 - r^2 ds_2^2
\end{equation}
with $f(r) = 1-2M/r$,
where $ds_2^2 = d\theta^2 + \sin^2\theta\,d\phi^2$ is the line element of
the unit two-sphere.  (The horizon radius is $r_H= 2M$.)
We compute the response rate of the static electric charge
to the Hawking radiation present in the Unruh vacuum.
The charge is 
placed at $(r,\theta,\phi) = (r_0,\theta_0,\phi_0)$ as described 
by the current density
\begin{equation}
j^\mu = \delta^\mu_{ t}\frac{q}{r^2\sin\theta}
\delta(r-r_0)\delta(\theta - \theta_0)
\delta(\phi-\phi_0) \,. \label{Jt}
\end{equation}
Since this current density is static, it couples to photons with zero
energy.  It turns out that the rate of response of this current to a single
photon vanishes.  However, the Bose-Einstein distribution of the thermal
photons coming out of the horizon diverges at zero energy.  This makes the
rates of absorption and stimulated emission indefinite.  This ambiguity can
be resolved by replacing the current density (\ref{Jt}) by
$j^\mu = (j^t, j^r, 0, 0)$, where
\begin{eqnarray}
j^t & = & \frac{\sqrt{2}\,q \cos Et}{r^2 \sin\theta}
\delta(r-r_0)\delta(\theta-\theta_0)\delta(\phi - \phi_0)\,, \\
j^r & = & \frac{\sqrt{2}q E\sin Et}{r^2 \sin\theta}
\Theta(r-r_0)\delta(\theta-\theta_0)\delta(\phi - \phi_0)\,. 
\end{eqnarray}
Here, $\Theta(x) = 1$ if $x>0$ and $\Theta(x) = 0$ if $x \leq 0$.
(The factor of $\sqrt{2}$ is necessary to make the time average of the
squared charge equal to $q^2$.)
We take the limit $E\to 0$ in the end,
assuming that the rate is continuous at $E=0$.
This continuity has been verified in 
Minkowski spacetime~\cite{HMS1}.

Since the current (\ref{Jt}) is conserved, it does not interact with 
pure-gauge particles
corresponding to 
the creation operators $a^{(3 n)\dagger}_{\omega lm}$. It does interact with
the states created by $a^{(0 n)\dagger}_{\omega lm}$ but these have
zero norm and do not contribute to physical probabilities.  Hence, as usual,
we need to consider only the physical modes.  Note also that only physical
modes I are relevant because $A_t=A_r =0$ for physical modes II.

Now, the transition amplitude from the Boulware vacuum $|0\rangle$ 
to the state 
$|1 n;\,\omega lm\rangle =a^{(1 n)\dagger}_{\omega l m}|0\rangle$ 
by the classical
current density $j^\mu$ is given by
\begin{equation}
{\cal A}^n_{\omega lm} = i\int d^4 x\sqrt{-g}j^\mu 
\langle 1 n ;\, \omega lm |\hat{A}_\mu|0\rangle
\,.  \label{Amp}
\end{equation}
Let us recast Eq.~(\ref{Amp}) as
\begin{equation}
{\cal A}^{n}_{\omega lm}\equiv 
2\pi i\, {\cal T}^{n}_{\omega lm}\delta(\omega - E)\,,
\label{Amp2}
\end{equation}
where ${\cal T}^{n}_{\omega lm}$ will be calculated later. The corresponding
transition rate, i.e. transition probability per asymptotic proper time,
is expressed as
\begin{equation}
{\cal R}^n_{\omega lm} = 2\pi |{\cal T}^{n}_{\omega lm}|^2
\delta(\omega - E)\,.
\end{equation}
Now we take into account that in
the Unruh vacuum there is a thermal flux of photons coming out
of the horizon with inverse temperature $\beta = 8\pi M$
(with no particles coming from infinity).  In the limit
$E\to 0$, only the absorption and stimulated emission of these thermal photons
contribute to the response rate.  For this reason, we treat only the modes 
coming out of the
horizon ($n = \rightarrow$).  The total response 
rate due to these modes with fixed
angular momentum, in which absorption, spontaneous and stimulated emissions
are all taken into account, is 
\begin{equation}
R_{Elm} = \int_0^{+\infty} d\omega {\cal R}^{\rightarrow}_{\omega lm}
\left( \frac{2}{e^{\beta\omega} - 1} + 1\right)
= 2\pi{\rm coth}\frac{\beta E}{2}|{\cal T}^{\rightarrow}_{E lm}|^2\,.
\end{equation}
Thus, in the limit $E\to 0$ we find
\begin{equation}
R_{0lm} = \lim_{E \to 0}
{|{\cal T}^{\rightarrow}_{E lm}|^2}/{(2ME)}\,.
\label{res}
\end{equation} 

Next ${\cal T}^{\rightarrow}_{E lm}$ will be calculated.
The equation satisfied by the function 
$q^{\rightarrow}_{\omega l}(r)\equiv (2M/r)
\varphi^{(1\rightarrow)}_{\omega l} (r)$
can be obtained from Eq.\ (\ref{varphi1}) with $p=2$ as
\begin{equation}
\frac{d\ }{dz} \left[(1-z^2)\frac{dq^{\rightarrow}_{\omega l}}{dz}\right]
+ \left[ l(l+1) - \frac{2}{z+1} - M^2\omega^2 \frac{(z+1)^3}{z-1}\right]
q^{\rightarrow}_{\omega l} = 0\,,  \label{equa}
\end{equation}
where $z  \equiv r/M - 1$.
For small $\omega$, the wave is almost completely reflected back to the 
horizon.  This implies that
\begin{equation}
\varphi^{(1\rightarrow)}_{\omega l} \approx 
- 2\sin [\omega (r^* -\tilde{r}_\omega)] 
\approx -2\omega r^* + {\rm const}.\;\;\;\;\; 
(r-r_H \ll \omega^2 r_H^3, \, |\omega r^*| \ll 1) \,,
\label{cond1}
\end{equation}
where $\tilde{r}_\omega$ are constants.  The
minus sign has been inserted for later convenience.
The Wheeler tortoise coordinate $r^*$ defined by Eq.\ (\ref{tortoise})
is written in Schwarzschild spacetime as 
\begin{equation}
r^* = r + 2M\ln\left({r \over 2M} - 1\right)\,.
\end{equation}
Thus, Eq.~(\ref{cond1}) can be written as 
\begin{equation}
\varphi^{(1\rightarrow)}_{\omega l}\approx
-4M\omega\ln(z-1)\, \;\;\;\;\;  
(r-r_H \ll \omega^2 r_H^3, \, |\omega r^*| \ll 1) \,. 
\label{cond2} 
\end{equation}
Now, Eq.\ (\ref{equa}) can be solved explicitly for $\omega = 0$.
By choosing the solution for $q^{\rightarrow}_{\omega l}$ such that
$\varphi^{(1\rightarrow)}_{\omega l} = [r/(2M)]q^{\rightarrow}_{\omega l} $ 
tends to zero as $z \to +\infty$, we find in the small $\omega$ limit
\begin{equation}
\varphi^{(1\rightarrow)}_{\omega l}
= 4M\omega (z+1)\left[ Q_l(z)-\frac{(z-1)}{l(l+1)}\frac{dQ_l(z)}{dz}\right]\,,
\label{cond3}
\end{equation}
where this has been normalized such that Eqs.~(\ref{cond2}) and (\ref{cond3})
are in agreement at $r\approx r_H$.

{}From gauge invariance of the amplitude (\ref{Amp}) it is clear that we
may use the modes in the spherical Coulomb gauge, 
$A^{(1'\rightarrow;\,\omega lm)}_\mu$, in place of those in the modified
Feynman gauge $A^{(1\rightarrow;\,\omega lm)}_\mu$.  Then the
contribution from the $r$-components will be suppressed in the low energy
limit due to extra factors of $\omega$ [see Eq.\ (\ref{suppr})].  Therefore
we need to consider only the $t$-component in this limit. 
The $t$-component can be found in the small $\omega$ limit 
from Eqs.\ (\ref{R1}),
(\ref{AtCoul}) and (\ref{cond3}) as 
\begin{equation}
A^{(1' \rightarrow;\,\omega lm)}_t
=  \frac{4i\omega(z-1)}{\sqrt{l(l+1)}}\,\frac{dQ_l(z)}{dz}
Y_{lm}(\theta,\phi) e^{-i\omega t}\,,
\label{A1'}
\end{equation}
where we have used the Legendre equation satisfied by $Q_l(z)$.
By using Eq.~(\ref{A1'}) in  Eq.~(\ref{Amp}) and comparing the result with
Eq.~(\ref{Amp2}), we obtain  
\begin{equation}
{\cal T}^{\rightarrow}_{\omega lm}=
-i\frac{\sqrt{2\omega}\,q(z_0 -1)}{\sqrt{\pi l(l+1)}}
\frac{dQ_l(z_0)}{dz_0}\overline Y_{lm}(\theta_0,\phi_0)\,.
\end{equation}
where $z_0 = r_0/M -1$.  Then we find the {\em proper} response rate 
of the charge as [see Eq.~(\ref{res})]
\begin{equation}
\frac{R_{0lm}}{\sqrt{f(r_0)}}
= \frac{q^2(z_0 -1)^2}{\pi M l(l+1)\sqrt{f(r_0)}}\left[
\frac{dQ_l(z_0)}{dz_0}\right]^2 |Y_{lm}(\theta_0,\phi_0)|^2\,,
\end{equation}
where the factor $\sqrt{f(r_0)}$ appears  because here we are calculating the
transition rate per proper time of the charge.
We can sum over $l$ and $m$ by using
\begin{equation}
\sum_{m=-l}^{l} |Y_{lm}(\theta_0,\phi_0)|^2
= {(2l+1)}/{(4\pi)}
\end{equation}
and
\begin{equation}
\sum_{l=1}^\infty \frac{2l+1}{l(l+1)}\left[ \frac{dQ_l(z)}{dz}\right]^2
= \frac{2Q_1(z)}{(z^2-1)^2}\,.  \label{QFORM}
\end{equation}
A derivation of this formula is given in Appendix B.
Thus, the total transition probability per proper time of the charge is
\begin{equation}
R^{\rm tot} = \sum_{l=1}^{+\infty} \sum_{m=-l}^{l}
\frac{R_{0lm}}{\sqrt{f(r_0)}}
= \frac{q^2 a(r_0) }{2\pi^2}Q_1(r_0/M-1)\,,
\end{equation}
where $a(r_0) = Mr_0^{-2}/\sqrt{f(r_0)}$ is the proper acceleration of the
charge.

It is of course possible to work with the modes in the modified Feynman gauge
directly and obtain the same result.  If we do so, the contribution will come
from the $r$-component of the current because the $t$-component of
physical modes I vanish. 

\section{Low energy absorption cross section of the Schwarzschild black hole}
\label{sec:Schwarz}

Here we consider the Schwarzschild black hole in $p+2$ dimensions given by
the line element~(\ref{ssle})
with $f(r) = 1- (r_H/r)^{p-1}$.  If the absorption probability of 
the black hole associated with physical modes I and II 
are ${\cal P}_l^{(1)}$ and ${\cal P}_l^{(2)}$, respectively, then the
absorption cross section is given by
\begin{equation}
\sigma = \frac{(2\pi)^p}{p\Omega_p \omega^p}\sum_{l=1}^\infty
\left[ M^{(1)}_l {\cal P}_l^{(1)} + M^{(2)}_l{\cal P}_l^{(2)}\right]\,,
\label{ABSCR}
\end{equation}
where $l$ and $\omega$ are the angular momentum and frequency, respectively,
of the modes and
$\Omega_p = 2\pi^{(p+1)/2}/\Gamma[(p+1)/2]$ is the volume of the unit
$p$-sphere.
Here, $M^{(1)}_l$ and $M^{(2)}_l$ are the multiplicities of the scalar and
divergence-free vector spherical
harmonics, respectively, on the $p$-sphere.  These multiplicities
are given by
(see, e.g., Refs.~\cite{CM,CH})
\begin{equation}
M^{(1)}_l = \frac{(2l+p-1)(l+p-2)!}{(p-1)!l!}\,,
\end{equation}
and
\begin{equation}
M^{(2)}_l = \frac{(2l+p-1)(l+p-1)!}{(l+1)(l+p-2)(p-2)!(l-1)!}\,.
\end{equation}
Eq.\ (\ref{ABSCR}) is derived in Appendix C.

For either physical modes I or II, we have for large $r (\approx r^*)$
that [see Eqs.~(\ref{varphi1}) and (\ref{varphi2})]
\begin{equation}
\left\{ \omega^2 + \frac{d^2\ }{dr^2} - \left[l(l+p-1)+\frac{p(p-2)}{4}\right]
\frac{1}{r^2}\right\}\varphi^{(\lambda\leftarrow)}_{\omega l}(r) \approx 0\,.
\end{equation}
The asymptotic solution with the correct normalization, 
i.e. with the coefficient of $e^{-i\omega r}$ being one, is
\begin{equation} 
\varphi^{(\lambda\leftarrow)}_{\omega l}(r)
\approx \sqrt{2\pi\omega r}J_{l+(p-1)/2}(\omega r)\,\ \  (r_H \ll r)\,.
\end{equation}
Thus, for small $\omega r$
\begin{equation}
\varphi^{(\lambda\leftarrow)}_{\omega l}(r)
\approx C_l \left( {r}/{r_H}\right)^{l+p/2} \,\ \ 
(r_H \ll r \ll \omega^{-1})\,, 
\label{Cl}
\end{equation}
where
\begin{equation}
C_l = \frac{\sqrt{4\pi}}{\Gamma\left( l+ (p+1)/2\right)}
\left(\frac{\omega r_H}{2}\right)^{l+p/2}\,.
\label{Clf}
\end{equation}

Now, near the horizon, where $V_\lambda \ll \omega^2\;\;$ ($\lambda=1,2$), 
these solutions behave like
[see Eqs.~(\ref{varphi1}) and (\ref{varphi2})]
\begin{equation}
\varphi^{(\lambda\leftarrow)}_{\omega l}(r)
\approx D_l^{(\lambda)}e^{-i\omega r^*}\,\ \ (r-r_H \ll \omega^2 r_H^3)\,,
\end{equation}
where the constants $D_l^{(\lambda)}$ will be determined later.
Hence, for $|\omega r^*| \ll 1$ we have in this region
$\varphi^{(\lambda\leftarrow)}_{\omega l}(r) \approx D_l^{(\lambda)}$ and the
absorption probability is given by 
${\cal P}_l^{(\lambda)} = |D_l^{(\lambda)}|^2$.

In order to find the coefficient $D_l^{(\lambda)}$, we solve 
the equations for
$\varphi^{(\lambda\leftarrow)}_{\omega l}(r)$ with $\omega = 0$.  For
physical modes I, it turns out that the function 
$r^p R^{(1\leftarrow)}_{\omega l}$ satisfies a hypergeometric equation
with the variable $w = 1- (r/r_H)^{p-1}$.  Thus, we find in the
small $\omega$ limit
\begin{equation}
\varphi^{(1\leftarrow)}_{\omega l}(r)
= D_l^{(1)}\left({r}/{r_H}\right)^{1-p/2}
F(-(l+p-1)/(p-1), l/(p-1);1;w)\,.
\label{F1}
\end{equation}
This function approaches $D_l^{(1)}$ for $r \to r_H$ as required.
By using the formula~\cite{GRAS}
\begin{equation}
F(\alpha,\beta;\gamma;w) \approx
\frac{\Gamma(\gamma)\Gamma(\beta-\alpha)}{\Gamma(\beta)\Gamma(\gamma-\alpha)}
(-w)^{-\alpha}\,\ \ (\alpha < \beta\,,\ \ -w\gg 1) \label{Hyper1}
\end{equation}
in Eq.~(\ref{F1}), we find asymptotically  
\begin{equation}
\varphi^{(1\leftarrow)}_{\omega l}(r)
\approx D_l^{(1)}E_l^{(1)}\left({r}/{r_H}\right)^{l+p/2}\, \ \
(r_H \ll r \ll \omega^{-1})\,, \label{DLEL}
\end{equation}
where
\begin{equation}
E_l^{(1)} = \frac{\Gamma\left[(2l+p-1)/(p-1)\right]}
{\Gamma\left[{l}/(p-1)\right]\Gamma\left[(l+2p-2)/(p-1)\right]}\,.
\end{equation}
By comparing Eqs.\ (\ref{Cl}) and (\ref{DLEL}), we obtain
\begin{equation}
{\cal P}^{(1)}_l = [C_l/E_l^{(1)}]^2 \,.
\label{P1S}
\end{equation}
Thus, the absorption probability behaves like $\omega^{2l+p}$.  

Similarly, for physical modes II, 
we find in the small $\omega$ limit
\begin{equation}
\varphi^{(2\leftarrow)}_{\omega l}(r)
= D_l^{(2)}\left({r}/{r_H}\right)^{(p/2 -1)}
F(-(l+1)/(p-1), (l+p-2)/(p-1);1;w)\,.
\end{equation}
Then by using Eq.\ (\ref{Hyper1}) we find asymptotically
\begin{equation}
\varphi^{(2\leftarrow)}_{\omega l}(r)
\approx D_l^{(2)}E_l^{(2)}\left({r}/{r_H}\right)^{l+p/2}\, \ \
(r_H \ll r \ll \omega^{-1}) \,, \label{DLEL2}
\end{equation}
where
\begin{equation}
E_l^{(2)} = \frac{\Gamma\left[(2l+p-1)/(p-1)\right]}
{\Gamma\left[(l+p-2)/(p-1)\right]\Gamma\left[(l+p)/(p-1)\right]}\,.
\end{equation}
By comparing Eqs.\ (\ref{Cl}) and (\ref{DLEL2}), we obtain 
\begin{equation}
{\cal P}^{(2)}_l = [C_l/E_l^{(2)}]^2 \,.
\label{P2S}
\end{equation}
Again, the absorption probability behaves like $\omega^{2l+p}$.
Therefore the physical modes with 
$l=1$ give the dominant contribution to the absorption probability,
as expected.

By substituting ${\cal P}^{(1)}_1$ and ${\cal P}^{(2)}_1$ in Eq.~(\ref{ABSCR}), 
we obtain the total absorption
cross section for low energy as
\begin{equation}
\sigma = \frac{\omega^2 r_H^2 A_H}{2(p+1)}\left[
\frac{2p\left[\Gamma\left({p}/(p-1)\right)\right]^4}
{\left[ \Gamma\left((p+1)/(p-1)\right)\right]^2} + 1\right]\,,
\end{equation}
where $A_H = \Omega_p r_H^p$ is the area of the black hole.  

One would need to solve the differential equations satisfied by
$\varphi^{(\lambda\leftarrow)}_{\omega l}$ numerically to find the 
low-energy absorption
cross section for non-extreme Reissner-Nordstr\"om black holes.\footnote{It
is possible to obtain a result in a closed form if one takes the limit
$\omega \to 0$ and $r_+ \to r_-$ simultaneously~\cite{Gubser}.}  However,
it is possible to calculate it analytically for the 
extreme Reissner-Nordstr\"om
black hole.  We turn to this problem in the next section.

\section{Low energy absorption cross section of the extreme 
Reissner-Nordstr\"om black hole}
\label{sec:Reissner}

The line element of the extreme Reissner-Nordstr\"om spacetime is given by
Eq.~(\ref{ssle}) with $f(r) = h(r)^{-1} = [1- (r_H/r)^{p-1}]^2$.  
As in the Schwarzschild case, we have
$\varphi^{(\lambda \leftarrow)}_{\omega l}(r)
\approx C_l (r/r_H)^{l+p/2}$ for $r_H \ll r\ll \omega^{-1}$
where $C_l$ is given by Eq.~(\ref{Clf}).

Now, let us analyze the physical modes close to the horizon.
Concerning  mode I, we write
\begin{equation}
\left[\omega^2 + \frac{d^2\ }{dr^{* 2}} - 
\frac{l(l+p-1)}{(p-1)^2r^{* 2}}\right]
\varphi^{(1n)}_{\omega l} \approx  0 \, \ \
\left( -\omega^2 {r^*}^3 \gg r_H \right)\, , 
\label{fi1}
\end{equation}
where we have used that the Wheeler tortoise coordinate
$r^*$ [see Eq.~(\ref{tortoise})] behaves in this region as
\begin{equation}
r^* \approx - \frac{(p-1)^{-2} r_H^2}{r-r_H} \,.
\label{r*}
\end{equation}
Hence we obtain 
\begin{equation}
\varphi^{(1\leftarrow)}_{\omega l}
\approx D_l^{(1)}\sqrt{\frac{-\pi\omega r^*}{2}}H_{\nu}^{(1)}(-\omega r^*)\,\ \
\left( -\omega^2 {r^*}^3 \gg r_H \right)
\label{fi1sol}
\end{equation}
with
\begin{equation}
\nu = \left[ \frac{1}{4} + \frac{l(l+p-1)}{(p-1)^2}\right]^{1/2}
= \frac{1}{2} + \frac{l}{p-1}\,.
\label{nu}
\end{equation}
This solution behaves  
like $D_l^{(1)}e^{-i\omega r^*}$ (up to a phase factor) very close to the
horizon ($-\omega r^* \gg 1$),
where the effective potential $V_1$ becomes negligible, as expected. Hence, the
absorption probability is ${\cal P}_l^{(1)} = |D_l^{(1)}|^2$.
Note that, for small $z$ with positive $\nu$, one has
\begin{equation}
H_\nu^{(1)}(z)\approx -\frac{i\Gamma(\nu)}{\pi}\left( \frac{2}{z}\right)^\nu\,.
\end{equation}
Hence,
\begin{equation}
\varphi^{(1\leftarrow)}_{\omega l}
\approx D_l^{(1)}K_l^{(1)}\left[ \frac{(p-1)(r-r_H)}{r_H}\right]^{\nu-1/2}\,
\ \ \left( (\omega r_H)^{1/3} \ll -\omega r^* \ll 1 \right)
\label{aux1RN}
\,,
\end{equation}
where
\begin{equation}
K_l^{(1)} =
- \frac{i\Gamma(\nu)}{\sqrt{\pi}}
\left[ \frac{2(p-1)}{\omega r_H}\right]^{\nu - 1/2}\,.
\end{equation}
Again, in order to determine $D_l^{(1)}$,   we solve 
the equation for
$\varphi^{(1 \leftarrow)}_{\omega l}(r)$ in the small $\omega$ limit:
\begin{equation}
\varphi^{(1\leftarrow)}_{\omega l}(r)
= D_l^{(1)} K_l^{(1)}\left( \frac{r}{r_H}\right)^{1-p/2}
\left(1 - \frac{l}{l+p-1}w\right)(-w)^{l/(p-1)}\,,
\label{aux2RN}
\end{equation} 
where $w = 1 - (r/r_H)^{p-1}$, and we note that Eq.~(\ref{aux2RN})
has been normalized such that it agrees with 
Eq.~(\ref{aux1RN}) near the horizon.
On the other hand, we find asymptotically 
\begin{equation}
\varphi^{(1\leftarrow)}_{\omega l}
\approx \frac{l}{l+p-1}
D^{(1)}_l K_l^{(1)}\left( \frac{r}{r_H}\right)^{l+p/2}\, \ \ 
(r_H \ll r \ll \omega^{-1})\,.
\end{equation}
By comparing this with Eq.\ (\ref{Cl}) we find
\begin{equation}
{\cal P}_l^{(1)} = \left|\frac{(l+p-1)C_l}{l \, K_l^{(1)}}\right|^2\,.
\label{Probl1}
\end{equation}
We note that ${\cal P}_l^{(1)} \propto \omega^{p+2lp/(p-1)}$.  Hence the 
dominant contribution to the scattering cross section comes from
the $l=1$ term:  
\begin{equation}
{\cal P}^{(1)}_1
= \frac{4\pi^2 p^2}
{\left[(p-1)^{1/(p-1)}
\Gamma\left[(p+3)/2 \right]\Gamma\left[(p+1)/(2p-2)\right]\right]^2}
\left( \frac{\omega r_H}{2}\right)^{2+p+2/(p-1)}\,.
\end{equation}
The contribution from physical modes II can be found in a
similar manner. By substituting Eq.~(\ref{r*}) in Eq.~(\ref{varphi2}),
we have near the horizon  
\begin{equation}
\left[\omega^2 + \frac{d^2\ }{dr^{* 2}} - 
\frac{(l+1)(l+p-2)}{(p-1)^2r^{* 2}}\right]
\varphi^{(2n)}_{\omega l} \approx  0 \, \ \ 
\left( -\omega^2 {r^*}^3 \gg r_H  \right) \,.  
\label{fi2}
\end{equation}
Then $\varphi^{(2 \leftarrow)}_{\omega l}$ will have the same form as 
$\varphi^{(1 \leftarrow)}_{\omega l}$ in Eq.~(\ref{fi1sol})
except that $\nu$ in Eq.~(\ref{nu}) is replaced by
$$
\nu' = \left[ \frac{1}{4} + \frac{(l+1)(l+p-2)}{(p-1)^2}\right]^{1/2}\,.
$$
For $p\geq 3$, the contribution from physical mode~II is suppressed
compared to that from physical modes~I since $\nu' > \nu$.  
For $p=2$, i.e. in four dimensions,
Eqs.~(\ref{fi1}) and (\ref{fi2}) satisfied by 
$\varphi^{(1n)}_{\omega l}$ and
$\varphi^{(2n)}_{\omega l}$ are the same.  Thus, the absorption
probabilities of these two types of modes are equal for each $l$.

Hence, we obtain the following results:
\begin{equation}
\sigma  =  \frac{4}{3} (\omega r_H)^4 A_H\,,\ \ {\rm for}\ p=2\,,
\label{gub}
\end{equation}
\begin{equation}
\sigma  =  \frac{\pi p(\omega r_H)^{2+2/(p-1)}A_H}
{(p+1)\left[ 2(p-1)\right]^{2/(p-1)}\left[\Gamma\left[{(p+1)}/{(2p-2)}\right]
\right]^2}\,,\ \ {\rm for}\ p \geq 3\,.
\end{equation}
Eq.\ (\ref{gub}) is in agreement with the result derived by 
Gubser~\cite{Gubser} .
It is interesting to note that the low energy absorption cross section
is proportional to a fractional power of $\omega$ if $p \geq 4$.

\section{Conclusions}
\label{sec:Conclusions}

In this paper we showed that the field equations for free electrodynamics
can be reduced to decoupled scalar field equations in a modified
Feynman gauge in the spacetime of spherically symmetric black hole.  
(We also noted that the equations for the physical modes
simplify in the spherical Coulomb gauge.)  Then we examined the Gupta-Bleuler
quantization in this modified Feynman gauge.

Next we reviewed the calculation of the response rate of a static charge
outside the four-dimensional Schwarzschild black hole. 
It is easy to extend our result to the Schwarzschild
black hole in arbitrary dimensions, though we cannot simplify the resulting 
infinite series. 
A closed-form expression for the response rate of a static source 
has been obtained for massless scalar field
in four-dimensional Reissner-Nordstr\"om
black hole~\cite{CaMa}.
The extension of the results in this paper 
to the Reissner-Nordstr\"om black hole
will also be possible in principle, 
but the result will probably 
not be expressible (even) as an infinite series of familiar special functions
in any dimensions.

Finally we calculated the absorption cross sections of low energy photons
by the Schwarzschild and extreme Reissner-Nordstr\"om black holes in
arbitrary dimensions.  The corresponding cross section of massless
scalar particles is known to be equal to
the horizon area~\cite{DGM} as long as the
black hole is spherically symmetric.  No such universality holds for photons
as our results and previous results~\cite{Gubser} show.  It is interesting 
that there are two modes, modes I and II, which behave differently in
higher dimensions.  It is also intriguing that the absorption cross section
scales as a fractional power of the energy $\omega$, i.e. as 
$\omega^{2+p+2/(p-1)}$ in $p+2$ dimensions.
Finally, it would be 
interesting to compare the
results here or similar results obtained using our gauge with 
those in string theory. (See, e.g., Ref.~\cite{Gubser} and references therein
for examples of such comparison.)

    
\acknowledgments
LC and GM would like to acknowledge partial financial 
support from CAPES through the PICDT program
and Conselho Nacional de Desenvolvimento Cient\'\i fico e
Tecnol\'ogico, respectively.

\newpage

\appendix
\section{Field-strength two-point function in spherical polar coordinates
in four-dimensional Minkowski spacetime}
\label{app:Mink}
Minkowski spacetime is the simplest spherically symmetric spacetime.
In this Appendix we compute some components of the two-point function 
\begin{equation}
G_{\mu\nu\rho\sigma}(x,x')
\equiv \langle 0|\hat{F}_{\mu\nu}(x)\hat{F}_{\rho\sigma}(x')|0\rangle
\label{GFF}
\end{equation}
in spherical polar coordinates and verify that they agree with those
previously computed in Cartesian coordinates in four-dimensional Minkowski
spacetime.
We need to consider only 
physical modes 
because pure-gauge modes do not contribute to $\hat{F}_{\mu\nu}$
and the coefficient operators of the
non-physical modes have zero commutators with all operators appearing on
the right-hand side of Eq.\ (\ref{GFF}).

Since there is no horizon, all modes come in from infinity and go out to
infinity. Therefore, we do not need here the label $n =\leftarrow \, ,\, 
\rightarrow$ which
distinguished between the modes coming out of the horizon and those incoming
from infinity.  By the remark after Eq.\ (\ref{normal1}) we have
$\varphi^{(1)}_{\omega l}(r) = 2\omega 
r j_l(\omega r)$,
where
$j_l(x)$ is the spherical Bessel function of order $l$, in four-dimensional
Minkowski spacetime. Hence for physical modes I we have exactly
\begin{equation}
A^{(1;\,\omega lm)}_r = \frac{2\sqrt{l(l+1)}}{r}
j_l(\omega r)Y_{lm}(\theta,\phi)
e^{-i\omega t}\,,
\end{equation}
where $Y_{lm}(\theta,\phi)$ are the familiar scalar spherical harmonics 
on $S^2$. The $tr$-component of the corresponding field strength is
\begin{equation}
F_{tr}^{(1;\,\omega lm)}
= -i\frac{2\sqrt{l(l+1)}\, \omega}{r}j_l(\omega r)Y_{lm}(\theta,\phi)
e^{-i\omega t}\,,
\end{equation}
since the $t$-component vanishes.  These modes are sufficient for
computing $G_{trtr}(x,x')$
because physical modes II have $A_t = A_r = 0$.
We obtain with $x=(t,r,\theta,\phi)$ and $x'=(t',r',\theta',\phi')$
\begin{eqnarray}
G_{t r t r} (x, x') &=& 
\frac{1}{\pi rr'} \sum_{l=1}^{\infty} l(l+1) 
\sum_{m=-l}^{l} Y_{l m} (\theta ,\phi) \overline Y_{l m} (\theta ' ,\phi ') 
\int_{0}^{\infty} d\omega \omega 
j_l(\omega r) j_l(\omega r') e^{-i\omega (t-t')} 
\nonumber\\
&=& \frac{1}{8 \pi^2 (rr')^2} \sum_{l=1}^{\infty} l(l+1)(2l+1) 
P_l (\cos{\gamma})
Q_l \left[ \frac{-(t-t'- i \epsilon)^2 +r^2 +r^{' 2}}{2rr'} \right] \, ,
\label{GtrSpher}
\end{eqnarray}
where we have used in the last step the formulas
$$
\sum_{m=-l}^{l} Y_{l m} (\theta ,\phi) \overline Y_{l m}  (\theta ' ,\phi ')
= \frac{2l+1}{4\pi} P_l (\cos{\gamma}) \, ,
$$
with
$\cos{\gamma} = \cos{\theta} \cos{\theta '} + 
\sin{\theta} \sin{\theta '} \cos{(\phi - \phi ')}$,
and \cite{GRAS}
$$
\int_{0}^{\infty} d\omega \omega
j_l(\omega r) j_l(\omega r') e^{-i\omega (t-t')} =
\frac{1}{2rr'} Q_l \left[ \frac{-(t-t'- i \epsilon)^2 
+r^2 +r^{' 2}}{2rr'} \right] \, .
$$
We will simplify Eq.\ (\ref{GtrSpher}) later.

Next we consider the component 
$$
G_{\theta\phi\theta\phi} (x,x')= 
\frac{\sin\theta\sin\theta'}{4}\tilde{\epsilon}^{ij}\tilde{\epsilon}^{i'j'}
G_{iji'j'}(x,x') \,,
$$
where the anti-symmetric tensor on $S^2$, $\tilde{\epsilon}^{ij}$, has
the component $\tilde{\epsilon}^{\theta\phi} = (\sin\theta)^{-1}$.
(This definition differs from that of Regge and Wheeler~\cite{RW} by
a minus sign.)
Only the modes $A^{(2;\,\omega lm)}_\mu$ contribute here. First, we have
$\varphi^{(2)}_{\omega l}(r) = 2\omega
r j_l(\omega r)$.
Then since $R^{(2)}_{\omega l}(r) = \varphi^{(2)}_{\omega l}(r)$ for $p=2$, 
we find
\begin{equation}
A^{(2;\,\omega lm)}_i = 2\omega r j_l(\omega r)Y_i^{(lm)}(\theta,\phi)
e^{-i\omega t}\,,
\end{equation}
where the transverse vector spherical harmonics $Y_i^{(lm)}$ are given by
\begin{equation}
Y_i^{(lm)} = \frac{ \tilde{\epsilon}_{ij}\tilde{\nabla}^j
Y_{lm}}{\sqrt{l(l+1)}}\,.
\end{equation}
The corresponding field strength is given by
\begin{equation}
\frac{1}{2}\tilde{\epsilon}^{ij}F_{ij}^{(2;\,\omega lm)}
= 2\sqrt{l(l+1)}\omega r j_l(\omega r)Y_{lm}(\theta,\phi)
e^{-i\omega t}\,.
\end{equation}
Then in exactly the same way as for $G_{trtr}$, we find
\begin{equation}
G_{\theta \phi \theta \phi} (x, x')
= \frac{\sin\theta\sin\theta'}{8 \pi ^2} 
\sum_{l=1}^{\infty} l(l+1)(2l+1) P_l (\cos{\gamma})
Q_l \left[ \frac{-(t-t'- i \epsilon)^2 +r^2 +r^{' 2}}{2rr'} \right] \, .
\label{GtefiSpher}
\end{equation}

On the other hand one can calculate
$G_{\alpha \beta \gamma \delta} (x, x')$ using plane
wave modes. In the Cartesian coordinate system this is given in components
as (see, e.g., Ref.~\cite{T})
\begin{equation}
G^{(C)}_{tata} (x, x') = -4\pi \{1+ 2[(\zeta _a)^2 - (\zeta _t)^2 ]
\xi ^{-2} \} G_6^+ (x, x') \, ,
\label{Gtata}
\end{equation}
\begin{equation}
G^{(C)}_{tatb} (x, x') = -8\pi \zeta _a \zeta _b \xi ^{-2} 
G_6^+ (x, x') \, ,
\label{Gtatb}
\end{equation}
\begin{equation}
G^{(C)}_{taab} (x, x') = G^{(C)}_{abta} (x, x') = 
-8\pi \zeta _t \zeta _b \xi ^{-2} 
G_6^+ (x, x') \, ,
\label{Gtaab}
\end{equation}
\begin{equation}
G^{(C)}_{abab} (x, x') = 4\pi \{1+ 2[(\zeta _a)^2 + (\zeta _b)^2 ]
\xi ^{-2} \} G_6^+ (x, x') \, ,
\label{Gabab}
\end{equation}
\begin{equation}
G^{(C)}_{acbc} (x, x') = 8\pi \zeta _a \zeta _b \xi ^{-2} 
G_6^+ (x, x') \, ,
\label{Gacbc}
\end{equation}
where $a,b,c$ take values $x,y,z$ with $a$, $b$ and $c$ being all
distinct 
(no summation convention is used here)  
and $\zeta ^{\alpha} = x^{\alpha} - x^{' \alpha}$.
We have defined
$G_6^+ (x, x') = (4\pi ^3 \xi ^4)^{-1}  \,$ ,
where
\begin{eqnarray}
\xi^2 & = & (t - t'-i\epsilon)^2 - (x - x')^2 
- (y - y')^2 - (z - z')^2\, \nonumber \\
& = & (t-t'-i\epsilon)^2 - r^2 - r^{\prime 2} + 2rr'\cos\gamma\,.
\end{eqnarray} 
The components $G_{t r t r}$ in spherical polar coordinates 
can be obtained from the Cartesian ones 
(\ref{Gtata})-(\ref{Gacbc})
by using standard tensor transformation:
\begin{eqnarray}
G_{t r t r} (x, x') &=& \frac{4\pi}{r r'}
\left( (xx' + yy' + zz') \xi ^2 +
2 \left[ (xy' - x' y)^2 + (xz' - x' z)^2 
\right. \right. 
\nonumber\\
& & \left. \left. + 
(yz' - y' z)^2 \right]
\right) \frac{G_6^+ (x, x')}{\xi ^2} \, , \nonumber \\
& = &
\frac{1}{\pi^2\xi^6}\left\{ \left[ (t-t'-i\epsilon)^2 -r^2 -r^{\prime 2}\right]
\cos\gamma + 2rr'\right\}\,. 
\label{GtrCart}
\end{eqnarray}
Similarly, we find
\begin{equation}
G_{\theta \phi \theta \phi} (x, x') = 
\frac{r^2 r^{\prime 2} \sin\theta \sin\theta'}{\pi^2\xi^6}
\left\{ \left[ (t-t'-i\epsilon)^2 -r^2 -r^{\prime 2}\right]
\cos\gamma + 2rr'\right\}\,. 
\label{GtefiCart}
\end{equation}
Eqs.\ (\ref{GtrCart}) and (\ref{GtrSpher}) 
and Eqs.\ (\ref{GtefiCart}) and (\ref{GtefiSpher}) can be shown to agree
by using the formula
\begin{equation}
\sum_{l=1}^\infty l(l+1)(2l+1)P_l(t)Q_l(z)
= \frac{2(tz - 1)}{(z-t)^3}\,,
\end{equation}
which can be proved by applying $(d/dt)[(t^2-1)d/dt]$ on both sides of
the well-known formula
\begin{equation}
\sum_{l=0}^\infty (2l+1)P_l(t)Q_l(z) = {1}/{(z-t)} \label{PLQL}
\end{equation}
and using the Legendre equation satisfied by $P_l(t)$.

\section{A derivation of Eq.\ (\ref{QFORM})}
\label{app:Formula}
Define
\begin{equation}
F(z) \equiv 
\sum_{l=1}^{\infty}\frac{2l + 1}{l(l+1)}\left[
\frac{d\ }{dz}Q_l(z)\right]^2\,.
\end{equation}
By the Legendre equation satisfied by $Q_l(z)$, we have 
\begin{equation}
\frac{d\ }{dz}\left\{ (z^2 -1)^2\left[\frac{d\ }{dz}Q_l(z)\right]^2\right\}
= 2l(l+1)(z^2 -1)Q_l(z)\frac{d\ }{dz}Q_l(z)\,.
\end{equation}
Hence,
\begin{equation}
\frac{d\ }{dz}\left[ (z^2 -1)^2F(z)\right]
= (z^2 -1)\frac{d\ }{dz}\left\{ \sum_{l=1}^{\infty}(2l+1)[Q_{l}(z)]^2\right\}\,.
\label{App2_0}
\end{equation}
Now recall that~\cite{HMS3}
\begin{equation}
\sum_{l=0}^\infty (2l+1)[Q_l(z)]^2 = \frac{1}{z^2 -1}\,. \label{App2_1}
\end{equation}
[This can be obtained by squaring both sides of Eq.\ (\ref{PLQL})
and integrating over the variable $t$ from $-1$ to $1$.]
By substituting Eq.\ (\ref{App2_1}) in Eq.\ (\ref{App2_0}), we obtain
\begin{equation}
\frac{d\ }{dz}\left[ (z^2 -1)^2F(z)\right]
= (z^2 -1)\frac{d\ }{dz}\left[ \frac{1}{z^2 -1} - 
\frac{1}{4}\left(\ln\frac{z+1}{z-1}\right)^2\right]
= -\frac{2z}{z^2 -1} + \ln\frac{z+1}{z-1}\,, \label{App2_2}
\end{equation}
where we have used
\begin{equation}
Q_0(z) = \frac{1}{2}\ln \frac{z+1}{z-1}\,.
\end{equation}
Integration of Eq.\ (\ref{App2_2}) leads to
\begin{equation}
(z^2 -1)^2 F(z) = z\ln\frac{z+1}{z-1} -2 + C\,,
\end{equation}
where $C$ is a constant.  Recalling that $Q_l(z) \sim z^{-l-1}$ as 
$z\to\infty$, we find that the left-hand side tends to zero in this limit.
Therefore $C=0$.
Hence,
\begin{equation}
F(z) = 
\frac{1}{(z^2 -1)^2}\left( z\ln \frac{z+1}{z-1} - 2 \right)= 
\frac{2Q_1(z)}{(z^2 -1)^2}\,.
\end{equation}

\section{Absorption cross section in spherical 
polar coordinates: A derivation of
Eq.\ (\ref{ABSCR})}
\label{app:Cross}

We consider the divergence-free vector eigenfunctions of the Laplacian on the 
$(p+1)$-dimensional Euclidean space.  That is, we examine the solutions
of the following equations: 
\begin{equation}
\nabla^b (\nabla_b B_a - \nabla_a B_b) = -\omega^2 B_a \label{EU1}
\end{equation}
and
\begin{equation}
\nabla^a B_a = 0\,. \label{EU2}
\end{equation}
Let ${\bf x} = (x_1,x_2,\cdots,x_{p+1})$ be the Cartesian coordinates.  
There are plane wave solutions of the form
\begin{equation}
B^{(C;\,\omega\hat{k}\hat{\epsilon})}_j = \hat{\epsilon}_j 
e^{i{\bf k}\cdot {\bf x}}
\end{equation}
with $\omega > 0$,
where $\hat{k}$ is a unit vector and ${\bf k} = \omega \hat{k}$, 
and $\hat{\epsilon}$ is a polarization unit vector
orthogonal to $\hat{k}$.  We define
\begin{equation}
(B^{(1)}, B^{(2)})_{\rm E} \equiv \int d^{p+1}{\bf x} 
\overline{B^{(1)}}_a B^{(2)a}\,,
\end{equation}
for any vectors $B^{(1)}_a$ and $B^{(2)}_a$ on this space.  Then
\begin{equation}
(B^{(C;\,\omega\hat{k}\hat{\epsilon})},
B^{(C;\,\omega'\hat{k}'\hat{\epsilon}')})_{\rm E}
= (2\pi)^{p+1}
\overline{\hat{\epsilon}}\cdot\hat{\epsilon}'\,\delta^{p+1}({\bf k}-{\bf k}')
\,.
\end{equation}

Now, suppose that $B_a^{(P;\,\omega \kappa)}$ are solutions of Eqs.\ (\ref{EU1})
and (\ref{EU2}) with a discrete label $\kappa$ satisfying
\begin{equation}
(B^{(P;\,\omega\kappa)}, B^{(P;\,\omega'\kappa')})_{\rm E} 
= 2\pi\,\delta_{\kappa\kappa'}\delta(\omega-\omega')\,. \label{normali}
\end{equation}
Suppose that the plane wave solutions are expanded as
\begin{equation}
B^{(C;\,\omega\hat{k}\hat{\epsilon})}_a
= \sum_\kappa \alpha(\omega,\hat{k},\hat{\epsilon};\,\kappa)
B_a^{(P;\,\omega \kappa)}\,,
\end{equation}
where $\alpha(\omega,\hat{k},\hat{\epsilon};\,\kappa)$ are constants.
As can be seen from the mode analysis in Section \ref{sec:GuptaBleuler},
a solution of Eqs.\ (\ref{EU1}) and (\ref{EU2}) 
satisfies the normalization condition (\ref{normali}) if its incoming
angular components 
at large $r$ are $B_i \sim r^{-p/2} e^{-i\omega r} Y_i$, 
with the spherical harmonic
on $p$-sphere $Y_i$ being normalized by
$\int d\Omega_p \overline{Y}^i Y_i = 1$. (The component $B_r$ can be shown
to be suppressed by an extra power of $r$.)  Thus, if we define the flux
so that the solution $B_a^{(C;\,\omega\hat{k}\hat{\epsilon})}$ has a unit
flux, then its content of the incoming mode 
$B_a^{(P;\,\omega \kappa)}$ is 
$|\alpha(\omega,\hat{k},\hat{\epsilon};\,\kappa)|^2$ per unit time. 

By using orthonormality of these solutions, we find
\begin{equation}
\alpha(\omega,\hat{k},\hat{\epsilon};\,\kappa)2\pi\,\delta(\omega-\omega')
=(B^{(P;\,\omega \kappa)},B^{(C;\,\omega'\hat{k}\hat{\epsilon})})_{\rm E}\,.
\end{equation}
{}From this one readily finds
\begin{equation}
|\alpha(\omega,\hat{k},\hat{\epsilon};\,\kappa)|^2 (2\pi)^2
\omega^p \delta(\omega-\omega'')
= 
\int_0^\infty d\omega'\,\omega'^p 
(B^{(P;\,\omega \kappa)}, B^{(C;\,\omega'\hat{k}\hat{\epsilon})})_{\rm E}
(B^{(C;\,\omega'\hat{k}\hat{\epsilon})},B^{(P;\,\omega''\kappa)})_{\rm E}\,.
\label{alpha1}
\end{equation}
Now, the average of
$\overline{\hat{\epsilon}}_je^{i\omega'\hat{k}\cdot{\bf x}} 
\hat{\epsilon}_l e^{-i\omega'\hat{k}\cdot{\bf x'}}$ over the polarization
vectors is 
$$
\frac{1}{p}
(\delta_{jl} - \hat{k}_j\hat{k}_l)e^{i\omega'\hat{k}\cdot({\bf x}-{\bf x}')}
= \frac{1}{p}
(\delta_{jl} - \frac{1}{\omega^2}\partial_j\partial'_l)
e^{i\omega'\hat{k}\cdot({\bf x}-{\bf x}')}\,,
$$
where the partial derivative $\partial_j$ is with respect to ${\bf x}$ whereas
the partial
derivative $\partial'_l$ is with respect to ${\bf x}'$.  Since
the modes
$B^{(P'\,\omega \kappa)}_a$ are divergence-free, these derivatives can be
dropped when we average Eq.\ (\ref{alpha1}) over the polarization vectors.
By averaging Eq.\ (\ref{alpha1}) 
over $\hat{\epsilon}$ and $\hat{k}$ and then using
\begin{equation}
\int d\omega'\,\omega^{\prime p} d\hat{k} e^{i\omega'\hat{k}
\cdot({\bf x}-{\bf x}')} = (2\pi)^p \delta^{p+1}({\bf x}-{\bf x}')\,,
\end{equation}
we find $|\alpha(\omega;\,\kappa)|^2$,
the average of $|\alpha(\omega,\hat{k},\hat{\epsilon};\,\kappa)|^2$
over $\hat{k}$ and $\hat{\epsilon}$, as
\begin{equation}
|\alpha(\omega;\,\kappa)|^2 \omega^p \delta(\omega-\omega'') = 
\frac{(2\pi)^{p-1}}{p\Omega_p}
(B^{(P;\,\omega \kappa)}, B^{(P,\omega'' \kappa)})_{\rm E}\,.
\end{equation}
Then from Eq.\ (\ref{normali}), we find
\begin{equation}
|\alpha(\omega;\,\kappa)|^2 =
\frac{(2\pi)^p}{p\Omega_p \omega^p}\,.  \label{fluxcontent}
\end{equation}
Thus, if a plane wave has a unit flux, it contains the incoming
mode with label $\kappa$ given  
by (\ref{fluxcontent}) per unit time on average.

The angular components $A_i$ of physical modes I
are gradients of scalar
spherical harmonics.
On the other hand, physical modes II are divergence-free vectors on the
$p$-sphere.  Let $M^{(1)}_l$ and $M_l^{(2)}$ be the multiplicities 
of these two types of modes with a given angular momentum $l$. Then the
number of physical modes I with angular momentum $l$ is $M^{(1)}_l$ and
that of physical modes II is $M^{(2)}_l$.  Then, if there is a plane wave
with a unit flux, the content of physical modes I and II with
angular momentum $l$ are $[(2\pi)^{p}/p\Omega_p\omega^p]M^{(1)}_l$ and
$[(2\pi)^{p}/p\Omega_p\omega^p]M^{(2)}_l$ per unit time, respectively.
Therefore,
if the absorption probabilities
of physical modes I and II by a spherical black hole are ${\cal P}_l^{(1)}$
and ${\cal P}_l^{(2)}$, respectively, then the absorption cross section
is given by Eq.\ (\ref{ABSCR}).

\end{document}